\documentclass[a4paper,conference]{IEEEtran}
\IEEEoverridecommandlockouts
\usepackage{amsmath,amssymb,amsfonts}
\usepackage{xcolor}
\usepackage{algorithmic}
\usepackage{algorithm}
\usepackage{array}
\usepackage[caption=false,font=normalsize,labelfont=sf,textfont=sf]{subfig}
\usepackage{textcomp}
\usepackage{stfloats}
\usepackage{url}
\usepackage{verbatim}
\usepackage{graphicx}
\usepackage{multirow}
\usepackage{cite}
\usepackage{booktabs}
\newcommand{\nop}[1]{}

\def\BibTeX{{\rm B\kern-.05em{\sc i\kern-.025em b}\kern-.08em
    T\kern-.1667em\lower.7ex\hbox{E}\kern-.125emX}}
    
\begin{document}

\title{A Versatility-Performance Balanced Hardware Architecture for Scene Text Detection}

\author{\IEEEauthorblockN{Yao Xin\textsuperscript{1}, Guoming Tang\textsuperscript{2}, Donglong Chen\textsuperscript{3*}, Rumin Zhang\textsuperscript{4}, Teng Liang\textsuperscript{1},\\ Ray C. C. Cheung\textsuperscript{5}~\IEEEmembership{, Senior Member,~IEEE}, \c{C}etin Kaya Ko\c{c}\textsuperscript{6}~\IEEEmembership{, Fellow,~IEEE}}
\IEEEauthorblockA{\textit{\textsuperscript{1}Peng Cheng Laboratory, Shenzhen, China,   
\textsuperscript{2}National University of Defense Technology, Changsha, China,}
\IEEEauthorblockA{\textsuperscript{3}BNU-HKBU United International College, Zhuhai, China,
\textsuperscript{4}Southern University of Science and Technology, Shenzhen, China,}
\IEEEauthorblockA{\textsuperscript{5}City University of Hong Kong, Hong Kong, China,
\textsuperscript{6}UC Santa Barbara, USA; NUAA, China; Iğdır University, Turkey}}

\thanks{\textsuperscript{*}Donglong Chen is the corresponding author: donglongchen@uic.edu.cn}}

\maketitle

\begin{abstract}
Detecting and extracting textual information from natural scene images needs Scene Text Detection (STD) algorithms. Fully Convolutional Neural Networks (FCNs) are usually utilized as the backbone model to extract features in these instance segmentation based STD algorithms. FCNs naturally come with high computational complexity. Furthermore, to keep up with the growing variety of models, flexible architectures are needed. In order to accelerate various STD algorithms efficiently, a versatility-performance balanced hardware architecture is proposed, together with a simple but efficient way of configuration. This architecture is able to compute different FCN models without hardware redesign. The optimization is focused on hardware with finely designed computing modules, while the versatility of different network reconfigurations is achieved by microcodes instead of a strenuously designed compiler. Multiple parallel techniques at different levels and several complexity-reduction methods are explored to speed up the FCN computation. Results from implementation show that, given the same tasks, the proposed system achieves a better throughput compared with the studied GPU. Particularly, our system reduces the comprehensive Operation Expense (OpEx) at GPU by 46\%, while the power efficiency is enhanced by 32\%. This work has been deployed in commercial applications and provided stable consumer text detection services.
\end{abstract}

\begin{IEEEkeywords}
Scene Text Detection (STD), Instance Segmentation, Fully Convolutional Neural Network (FCN), Filed-Programmable-Gate-Array (FPGA), Hardware Acceleration.
\end{IEEEkeywords}

\section{Introduction}
\label{sect:intro}
{S}{cene} text detection is widely used in consumer electronics applications to facilitate Optical Characteristic Recognition (OCR). Scene text refers to text appearing in camera captured outdoor images. The task to determine the geometric information (including position and shape) of scene text from the input images is named scene text detection (STD), which acts as an essential prerequisite for subsequent scene text recognition (like ID card scanning, vehicle license plate recognition, etc.). Nevertheless, the task of STD is challenging due to various factors causing the image degradation, e.g., out-of-shape fonts, transformed styles, and light/shadow occlusion~\cite{lin2020review}.
  
To this end, increasing efforts have been applied to improving the efficiency and accuracy of STD systems~\cite{Zhang2016, LSB+2017}. A majority of state-of-the-art techniques are based on deep learning and they are dependent on a step of bounding box regression. In these methods, three typical modules are found and performed widely: text/non-text classification, location regression, and other post-processing (e.g., non-maximum suppression or merging text segments)~\cite{LSB+2017,Tian2016}. Particularly, various algorithms of text/non-text classification were developed to achieve \emph{semantic segmentation}~\cite{Zhang2016,Shi_2017_CVPR}. Due to the fact that text instances in scene images usually lie very close to each other (as shown in~\cite{deng2018pixellink}), however, they cannot be used to separate different text-lines effectively.

To tackle the above problem, \emph{instance segmentation} based methods were proposed to conduct dense (pixel-level) predictions~\cite{deng2018pixellink,long2020scene}. These methods employed an end-to-end Fully Convolutional Neural Network (FCN) to classify each pixel in images as text or non-text by generating a dense prediction map. Then a post-process groups pixels belonging to the same text together. Text bounding boxes can thus be generated directly without any regression operations. During post-processing, how to separate instances belonging to one text from the others is non-trivial. For example, PixelLink~\cite{deng2018pixellink} uses convolution to produce predicted links for each pixel and learn to predict whether two adjacent pixels belong to the same text instance by checking the positiveness of links.

\subsection{Motivations and Challenges}

Although the FCN-based algorithm is more effective than the semantic segmentation based one, it comes with a higher computation cost (sometimes even several orders of magnitude larger). In real-world implementations, Wang et al. ~\cite{Wang2022} proposed a distributed cloud-edge computing model to tackle the large-scale data at the cloud while leaving the small-scale data at the edge, and it was demonstrated that the performance of computing system could be improved efficiently. In terms of processing device, general purpose Graphics Processing Units (GPUs) could be applied to accelerate the model training and inference~\cite{NVS+2017}. However, due to the high power consumption and purchase cost of GPUs, their applications in power and cost efficient situations are much limited. In a prevail cloud-edge computing system~\cite{Kang2022, DOU20221}, GPUs are only suitable for deployment in cloud server. The needs of the edge also need to be considered, such as low power consumption and low latency.
Under such circumstances, FPGA acceleration systems have become popular in data centers and edge systems, due to their properties of high energy efficiency, reconfigurability and short turn-around period. Thus, FPGA or FPGA alike hardware is well motivated to be attempted for the FCN-based STD in our situation.

\begin{table*}
\centering
\caption{Related works comparison}
\label{tab:compare}
\setlength{\tabcolsep}{1.7mm}{
\footnotesize
\begin{tabular}{|c|c|c|c|c|c|}
\hline
 & Application   & Algorithm  & Versatility  & Random size & High performance  \\
\hline
Oliveira et al.~\cite{Oliveira2018}    & STR & HOG \& ELM  & Not support & No & No\\
Zho et al.~\cite{Zho2016}     & CNY banknote recognition & Projection \& Small FCN & Not support & No & No\\ 
Jing et al.~\cite{Jing2017}      & License plate recognition & Feed forward neural network & Not support & No & No\\ 
Zhao et al.~\cite{Zhao2019}      & STR & Binary SegNet & Not support & No & Yes\\
Xin et al.~\cite{Xin2021}      & STD & RRPN & Not support & Yes & Yes\\
Our work        & STD & FCN & Support & Yes & Yes\\ 
\hline
\end{tabular}}
\vspace{-0.4cm} 
\end{table*}

There are already some references describing FPGA hardware architectures designed for FCN operations. According to their purpose, these architectures can be generally classified into two categories: dedicated architecture designs for high performance~\cite{LLS+2019,Chang2020}  and general-purpose designs for multi-model support~\cite{Ma2017,Zhang2018,Cheng2018,Xing2019,Meng2021}. Both have their own advantages and disadvantages.
\begin{itemize}
    \item The dedicated architectures usually concentrate on particular algorithms in order to achieve efficient computation. However, this method lacks versatility because if the algorithm changes or adjusts, the hardware architectures need to be redesigned.
    \item The general-purpose designs emerged to support a  variety of network algorithms, without the need of redesign of the architecture. However, the generalization leads to performance compromise, because complex control generally requires more logic resources. Complicated software compilation tools are also needed to realize optimization and scheduling.
\end{itemize}

Therefore, efficiently combining generalization and high-performance (which we call  \emph{versatility-performance balance}) is a real-world challenge for the FPGA hardware architecture design. Additionally, as the accumulation operation in the very multi-layer convolution of FCNs could cause exponentially expanding errors, how to ensure a satisfactory accuracy, especially under the constraint of limited hardware resource, is another critical problem to tackle.

\addtolength{\topmargin}{0.01in}

\subsection{Our Contributions}

In this work, we propose a flexible hardware architecture that is tailored for the instance segmentation based STD algorithms and achieves a good versatility-performance balance for consumer applications. Specifically, the main contributions of this work are summarized as follows:

\begin{enumerate}

\item In pursuit of versatility, a concise and efficient method is proposed to achieve hardware generalization by using microcode. Specifically, the backbone networks for feature extraction and feature merging are both configured in advance without any hardware modification. The configuration of each layer and the interconnection of different layers are realized through diversified microcode sequence combinations. The method is hardware-friendly without the need of extra resources to execute FCN models or to configure complicated optimization combinations. Moreover, a full-stack auto configuration software tool chain is proposed and developed to facilitate microcode generation and model parameter normalization.

\item In order to maintain high performance, multiple parallel techniques (at channel level, buffer level, and module level) are explored to speed up the FCN computation. An efficient method is applied to merge the batch norm layer into convolutional layer. Furthermore, a novel technique is proposed to minimize the padding, thus reducing the computing complexity of upsampling module by 75\%.

\item For accuracy, fine tuned Block floating-point (BFP) data representations and accuracy maintenance techniques are adopted to achieve the optimal point in the trade-off between hardware resources and algorithmic accuracy. The experimental results show that the proposed FPGA design delivers STD acceleration with a high cost efficiency and a high performance while the decrease of f-measure for the benchmark dataset is only 0.55\%. It outperforms the studied GPU in terms of inference throughput, while its operating expense is lower than GPU by 46\% and the power efficiency is enhanced by 32\%.

\end{enumerate}

The rest of this paper is organized as follows: Section~\ref{sect:related} overviews related studies of hardware designs for text detection and recognition. Section~\ref{sec:sys_design} introduces the STD algorithm and proposes the auto configuration heterogeneous system and software tool chain. The hardware architecture and its optimization techniques are described in Section~\ref{sec:hardware}. The implementation results, performance evaluation
and precision comparison are shown in Section~\ref{sec:exp_result}. Section~\ref{sec:cls} concludes this paper.

\section{Related Work}
\label{sect:related}
\subsection{Fully Convolutional Networks Architecture}

Intensive research has been conducted in the general-purpose hardware accelerators for FCNs. 
\cite{Ma2017} presented a scalable and modularized Convolutional Neural Network (CNN) FPGA accelerator for ResNets. A layer-by-layer computation flow is designed to integrate computing primitives and communicate the complex connections of deep ResNet layers. But only ResNet-50 and ResNet-152 are supported with the throughput of 285.1 GOPS and 315.5 respectively.  
\cite{Zhang2018} proposed an automated tool flow that can transform Deep Neural Network (DNN) designs from popular deep learning frameworks to highly optimized board-level FPGA implementations. It is mainly focusing on the resource allocation and memory bandwidth adjustment. The tool is demonstrated on four DNNs (Alexnet, ZF, VGG-16, and YOLO). 

Xing et al\cite{Xing2019} proposed a full-stack compiler solution which is an integration of optimizers for graphs, loops and data layouts, an assembler, a runtime supporter and a validation environment. In this compiler, the fusion and pipeline optimization are explored with a subgraph isomorphism algorithm and a shortest-path heuristic. It achieves a throughput of 2.82 TOPs/s and 1.38 TOPs/s for VGG and ResNet50 with the implementation on ZU9 @330 MHz.
In~Meng's research\cite{Meng2021}, an algorithm-architecture co-optimization framework, named DYNAMAP, was proposed, which consists of a unified hardware overlay that can be reused across layers, supporting dynamic mapping of all three families of popular convolution algorithms, and a novel software Design Space Exploration (DSE) flow that customizes the hardware overlay and chooses optimal strategy mapping. 
It is observation that the state-of-art versatile architecture designs rely on the complicated optimization software, and the mapping manner from algorithm to hardware has multiple parameters which can hardly achieve across-the-board optimization.

\addtolength{\topmargin}{0.05in}

\subsection{Optical Character Recognition Architecture}

In published hardware designs for STD, Scene Text Recognition (STR), and
Optical Character Recognition (OCR) applications mainly use traditional
methods, such as Discrete Wavelet Transform (DWT), Histogram of Oriented
Gradients (HOG), and Maximally Stable Extremal Region (MSER).

An FPGA-CPU heterogeneous system for embedded STR applications is proposed
in~\cite{Oliveira2018}. The system combines hardware and software to
accelerate STR, and uses HOG for feature extraction and deploys a
neural network extreme learning machine as a classifier. For task
division, the FPGA acts as a bicubic interpolation accelerator to resize 
input images of any size to the size of $128\times128$ pixels. Other tasks are
performed by an Intel Atom N2600 processor. The results show that the system
reaches a good trade-off between processing time and recognition accuracy in
embedded environments.

In terms of scene text recognition, a computing-in-memory accelerator using the
binary SegNet is developed~\cite{Zhao2019}. The accelerator performs highly
efficient pixel-wise character classification by maximizing the bit-level
parallelism as well as optimizing the pipeline for low latency at the critical
path. The FPGA implementation is able to process the STR with an
energy-efficiency of 351.7 GOPs/W and a throughput of 307 fps for image of size
$128\times32$ pixels.

Regarding OCR applications, hardware designs for Chinese banknote
recognition~\cite{Zho2016} and car plate recognition~\cite{Jing2017} are
proposed. For the former application, the proposed system contains two stages:
character segmentation (CS) stage and OCR stage. The CS stage is similar to the
STD process, which utilizes vertical/horizontal projection for character
segmentation, while the OCR stage uses a small FCN to recognize the segmented
characters. The latter design adopts a three-layer feed forward neural network
and gets a high recognition accuracy of 98.20\%. However, the algorithms in
these systems are only suitable for small size image
recognition.

The first full-stack hardware architecture design of STD is proposed in~\cite{Xin2021}. An FPGA-CPU heterogeneous system is designed to speed up the throughput and reduce the energy usage. They present a hardware/software partition method to analyze and split the detection tasks to enhance hardware efficiency. The Winograd algorithm are utilized to reduce multiplication complexity. Experimental results show that the throughput of their heterogeneous system achieves 40 times and 1.4 times improvements compared with CPU and GPU, respectively. However, due to the dedicated design methodology, their architecture is only able to compute Rotation Region Proposal Networks (RRPN) with VGG as the backbone. Any change on the parameters or backbone network would result in a time-consuming hardware redesign.

The comparison between related architectures and this work is summarized in
Table~\ref{tab:compare}. Though there are several hardware designs in OCR
related fields, they are either dedicated design for a fixed algorithm or they do not
support random size input. The design of a general-purpose,
high-performance and high-accuracy deep learning method based STD system that supports random size input will be a valuable reference for both research and industry.

\section{STD Algorithm \& Auto Configuration System Design}
\begin{figure}
\centering
\includegraphics[width=0.48\textwidth, scale=0.35,keepaspectratio]{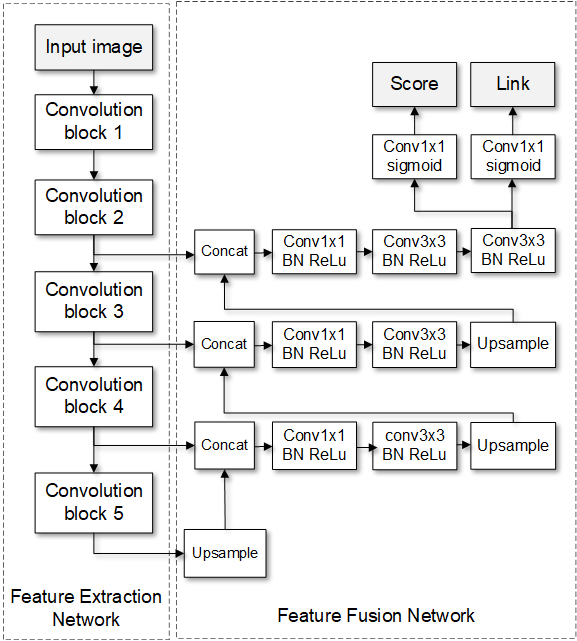}
\caption{The complete procedure of scene text detection algorithm.} 
\label{figure:std_fw}
\vspace{-0.48cm} 
\end{figure}

\label{sec:sys_design}

\subsection{STD Algorithm}
\label{sect:alg_overview}

We develop the scene text detection algorithm based on PixelLink~\cite{deng2018pixellink} and EAST~\cite{ZYW+2017}, the procedure of which is illustrated in Figure~\ref
{figure:std_fw}. More specifically, the feature extraction network and feature fusion network constitute a U-shape FCN. The feature extraction network is a convolutional network with interleaving convolution and pooling layers. Four levels of feature maps are extracted from the extractor network, whose sizes are 1/4, 1/8, 1/16, 1/32 of the input image, respectively. Then, these four feature maps are merged gradually through the fusion network. The technical rationale behind this is that multi-scale inception features are aggregated to encode rich local and context information for text prediction. The features from later stages of a neural network can determine large text geometry. While the low-level information in the early stages can determine small text geometry information. 

The final output makes two kinds of pixel-wise predictions which are text/non-text prediction (represented as score) and link prediction. The score determines if pixels are within text instances. Link prediction contains 8 elements for every pixel denoting 8 neighboring pixels. If the links between a given pixel and its neighbors are labeled as positive, it means they lie within the same instance. The positive score pixels are joined together into Connected Components (CC) according to predicted positive links, and each CC represents a detected text box. This is how instance segmentation is achieved without regression. 

The feature extraction network has several candidates such as ResNet, VGG, and MobileNet. Different network structures have distinct properties in terms of speed, accuracy, and training difficulty. Developers can select appropriate networks to meet specific requirements during deployment. To support a maximum degree of flexibility, the architecture of FCN is made to be capable of being configured by microcodes. Within the generally designed FCN architecture, the developer can modify the microcode to compute different networks. This architecture eliminates the tedious procedure of redesigning the whole hardware architecture when the network changes.

\addtolength{\topmargin}{0.05in}

\begin{figure}
\centering
\includegraphics[width=0.48\textwidth, scale=0.5,keepaspectratio]{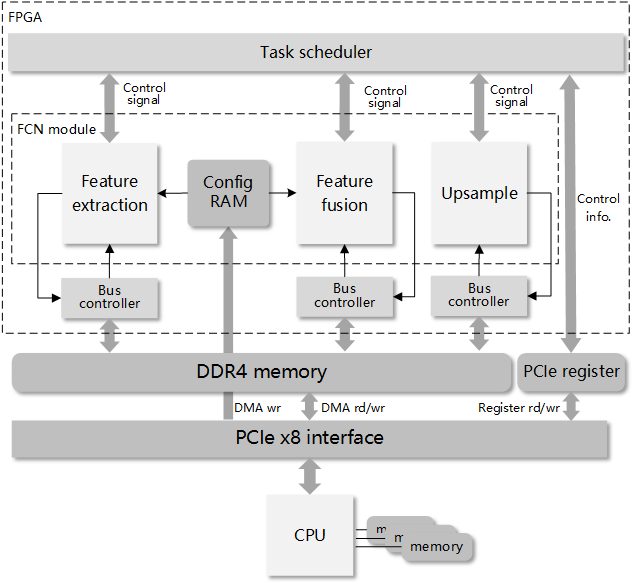}
\caption{Top-level architecture design of the proposed heterogeneous system.} 
\label{figure:top_arch}
\vspace{-0.48cm} 
\end{figure}

The top-level design of our heterogeneous system is shown in Figure~\ref{figure:top_arch}. The CPU acts as the control host of the whole system and the FPGA acts as a worker. The PCIe register accommodates control information and parameters for computing units, which can be accessed by both CPU and FPGA. According to the control information in the PCIe register, the task scheduler arranges the operation of different modules in a specific timing order. The processing data is accessed through the PCIe interface and is temporarily stored in DDR4 memory as a data pool. All computing modules read/write data from/to DDR4 via bus controllers.

In terms of FCN architecture, the feature extraction module and feature fusion module are responsible for the major processing tasks. This architecture facilitates a more flexible realization of different feature extraction and fusion networks according to the control of microcodes, compared with other
designs for specific purposes. The upsample module is independently extracted from the fusion module to make the architecture modularized in a coarse-grained manner, as the feature extraction, feature fusion and upsample can work independently in parallel 

The whole system works as follows. First, the configuration microcodes and model data are pre-loaded into DDR4 memories through PCIe interface from CPU host by DMA write. Then, CPU writes the related control registers of computing modules to invoke the module computation. The microcodes are then loaded into the configuration RAM and parsed successively. The FCN module is configured to load/store data and implement different types of layers according to the parsed parameters. After the above preparations, images are transferred to DDR4 memories from host CPU continuously. During the computation, the FCN module repeatedly reads the related data, computes according to the pre-determined dataflow, and writes the results back to memory. Under most circumstances, the results from the previous layer are treated as the input of the imminent layer. By keeping the temporary data of none-last layers in memory, the interactions between CPU and FPGA are reduced to a minimum level. Finally, the task scheduler issues an interrupt to the CPU when that round of computation is finished.

\begin{table*}
\setlength{\abovecaptionskip}{0cm}
\setlength{\belowcaptionskip}{1pt}
\scriptsize
\centering
\caption{Microcode Format}
\label{tab:microcode}
\setlength{\tabcolsep}{1.3mm}{
\begin{tabular}{ |c|c|c|c|c|c|c|c|c|c|c|c|c| }
\hline


 {Field} & Layer &{Transpose}& Input & Output & {Height} &{Width} & Kernel & {Stride} &{Res OP} & Input & Output &{Reserved}  \\
                    & type & \& Relu & channel & channel & & & size & & & addr & addr &  \\

\hline
Bitwidth       & 2 & 2 & 16 & 16 & 20 & 15 & 2 & 1 & 2 & 34 &34 & 112\\
\hline
\end{tabular}}
\vspace{-0.3cm} 
\end{table*}

\begin{figure*}
\setlength{\abovecaptionskip}{0cm}
\setlength{\belowcaptionskip}{1pt}
\centering
\includegraphics[width=0.85\textwidth, scale=0.45,keepaspectratio]{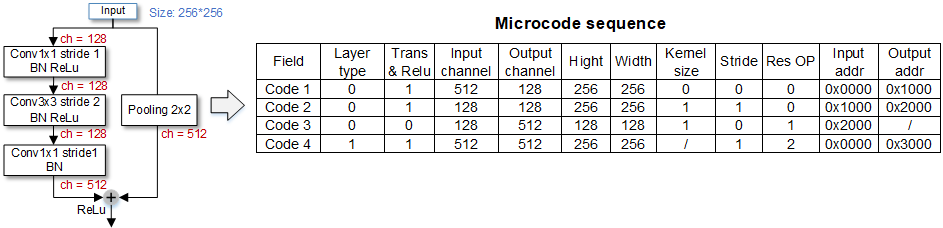}
\caption{An example of microcode sequences to compute a residual bottleneck in ResNet.} 
\label{figure:micro_ex}
\vspace{-0.4cm} 
\end{figure*}

\begin{figure}
\setlength{\abovecaptionskip}{0cm}
\setlength{\belowcaptionskip}{1pt}
\centering
\includegraphics[width=0.35\textwidth, scale=0.35,keepaspectratio]{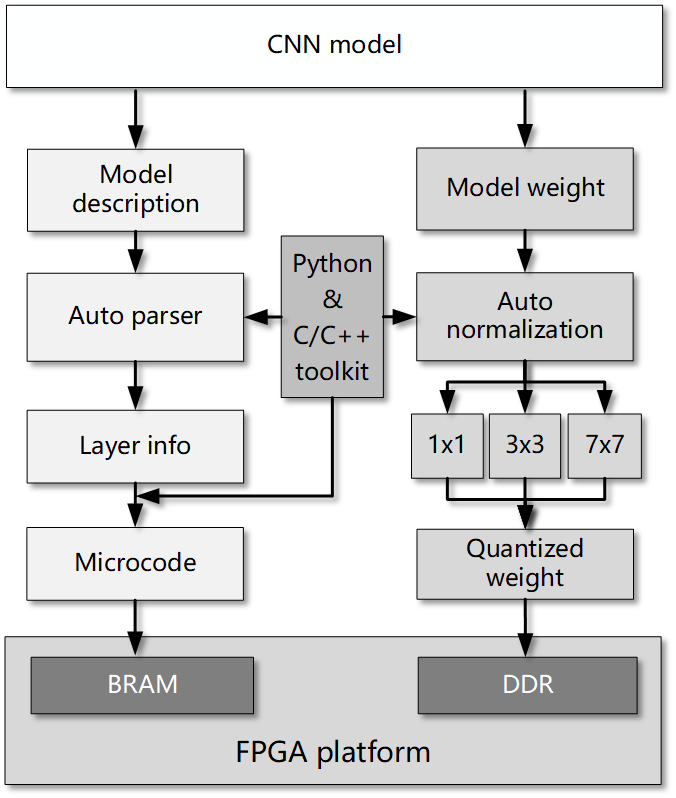}
\caption{The process flow of auto configuration. The left branch is FCN microcode generation and the right branch is model weight normalization.} 
\label{figure:auto-config}
\vspace{-0.5cm} 
\end{figure}

\subsection{Microcode Design}
\label{subsec:microcode}

Given that the increasing diversity and layer numbers of evolving FCN structures, how to design an efficient hardware is of great challenge. A relatively fixed computation dataflow can hardly accommodate varying FCN structures. In order to design a versatile hardware configuration scheme, we split the FCN structure into units of layers, and each FCN layer can be represented by a set of hyperparameters. Based on this observation, these hyperparameters of one layer are encoded into a microcode so that the configuration of the FCN networks is transformed to hyperparameters assignment, which is achieved by interpretation of a set of microcodes. 

The bit width of the microcode is 256-bit, which is aligned with the bit width of AXI data bus. One microcode is responsible for the hyperparameter setting of one specific layer. Table~\ref{tab:microcode} shows the format of a microcode. To be specific, layer type includes convolution, pooling, upsample, and null. Kernel size supports 1$\times$1, 3$\times$3, and 7$\times$7. Stride number supports 1 and 2. The residual OP is set specially for ResNet, and the value of 0 means no residual operation, 1 means layer results should be cached, 2 denotes layer results should be added to the cached result to get the final output of a residual block. The connection between layers is maintained by the input and output address allocation in external memory. Specifically, the results from each layer are stored in external memory as the input of the subsequent layer. The concatenation (shown as concat in Figure~\ref{figure:std_fw}) is achieved by putting the results of two layers into adjacent addresses. For ease of understanding, Figure~\ref{figure:micro_ex} provides an example of microcode sequences to compute a residual bottleneck in ResNet.

The complete set of microcodes is generated according to the FCN structure and transferred to the configuration on-chip RAM in the initialization phase. The microcode interpreter (shown in Figure~\ref{figure:fe_mod}) parses the microcode of each layer and distributes the parsed parameters to different units of FCN module. The FCN module works under the control of the parsed
parameters to perform the layer computation. 

\subsection{Auto Configuration Flow}
\label{subsec:conFigure}

The auto configuration process is shown in Figure~\ref{figure:auto-config}. It
has two branches: FCN microcode generation and model weight normalization.
Python and C/C++ toolkit have been developed to make the process highly
automated. The model description file is analyzed and resolved into the general
model description and further transformed to microcodes by a Python parser
layer by layer. The parser can resolve most types of FCN models with
convolution kernels of 1$\times$1, 3$\times$3, 7$\times$7, including residual
networks. Because block floating-point (BFP) format is applied in the
computation, the weights need to be normalized in advance. The BFP
normalization process is computed by the C/C++ toolkit. Exponent and mantissa
bitwidth are customized to obtain different precision combinations according
to the normalization block size and kernel size.

\subsection{Hardware Architecture Design of FCN Module}
\label{subsec:fcn_mod}

\begin{figure*}
\setlength{\abovecaptionskip}{0cm}
\setlength{\belowcaptionskip}{1pt}
\centering
\includegraphics[width=\textwidth, scale=0.5,keepaspectratio]{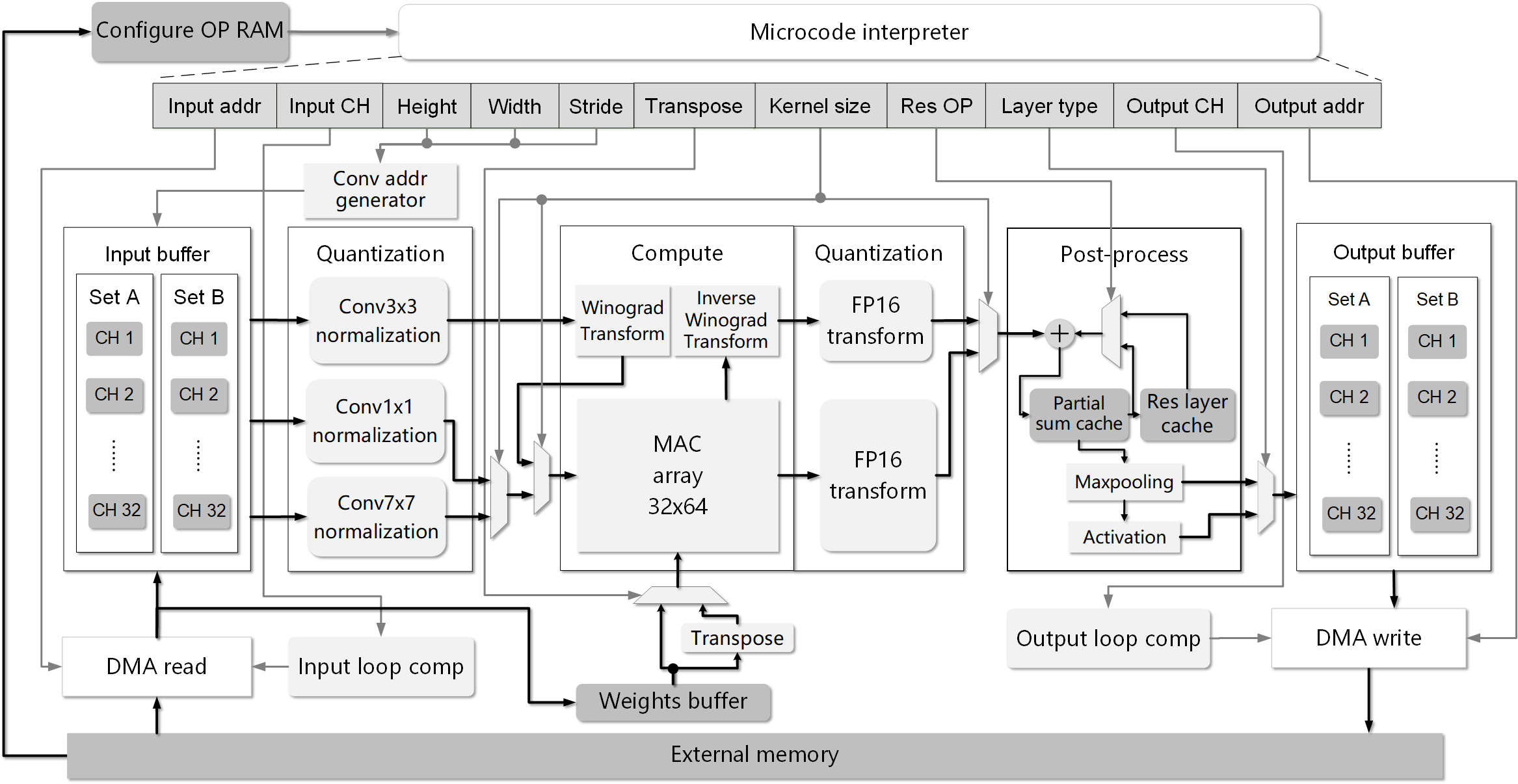}
\caption{The hardware architecture of feature extraction module.} 
\label{figure:fe_mod}
\vspace{-0.3cm} 
\end{figure*}

According to the property of the instance segmentation based STD algorithm, we divided the core computing section into two parts: feature extraction and feature fusion, which are all fully convolution networks with different network structures. Figure~\ref{figure:fe_mod} shows the hardware architecture of the feature extraction module. Details of microcode interpreter are also illustrated on how the parsed microcode controls the operation of each unit. Note that the feature fusion module is designed in a similar manner, but without a conv 7$\times$7 datapath, and max pooling is replaced by sigmoid. Moreover, the dimensions of the multiply-and-accumulate (MAC) arrays in these two FCN modules are different, which are 32$\times$64 and 16$\times$32, respectively.

The MAC arrays in the feature extraction module are built with DSP supertiles to perform fixed-point MAC on the mantissa part of BFP. The FCN module supports multiple size kernels of convolution: 1$\times$1, 3$\times$3, and 7$\times$7. These three types of convolutions share the same set of DSP arrays but have distinct input datapaths. Since 3$\times$3 convolution is dominant in various FCNs, Winograd algorithm with a tile size of $F(4\times4,3\times3)$ is deployed to reduce the number of multiplication by a factor of 4. The minimal algorithm for $F(4\times4,3\times3)$ can be formulated using the fixed
transformation matrices $A$, $B$ and $G$ as follows: 

\begin{equation}
Y = A^T[(GWG^T) \odot (B^TXB)]A
\end{equation}

\noindent where $\odot$ indicates point-wise multiplication, 
$W$ is a $3\times3$ kernel, $X$ is a $6\times6$ input image tile, and $Y$ is a $4\times4$ output. The number of multiplications for Winograd $F(4\times4,3\times3)$ is 36 compared to the number of 144 for conventional algorithm, which is a fourfold reduction. Moreover, given that the transformation matrices are fixed and weights are known, $GWG^T$ could be precomputed and stored for the subsequent operations.

The MAC array is implemented with DSP supertile arrays by using cascaded DSPs via two dimensions~\cite{Wu2017}. A supertile array consists of both memory and DSP. The memory in the array stores the pre-computed weights and works in a ping-pong mode thus the weights transfer time is overlapped with computation time. Note that Winograd input/output transforms are essentially matrix multiplication of the fixed matrices shown above, so one could rearrange the computation flow to reduce the hardware usage. Specifically, the input transform includes the multiplication of $B^T$ and input matrix $X$. 
A direct computation requires 12 multiplications and 16 add/sub operations. By rearranging the computation flow, it only requires 6 multiplications and 18 add/sub operations.

Convolutions with kernels of sizes 1$\times$1 and 7$\times$7 are performed
through the point-wise MAC method. These two datapaths bypass the Winograd
transform and lead to MAC arrays directly. Max pooling operation is embedded in
post-processing module, which could hide the computation time in a pipelined
manner.

We set the input dimension $M = 32$ and output dimension $N = 64$ for MAC
arrays in this design. The working clock frequency of the DSP array is twice
the clock frequency of the outside input/output interface. Thus doubling the
bitwidth of the input/output interface could support the data feeding of the inner
computing logics. With the help of the ping-pong mode of the input buffer, the MAC
is able to work at full capacity. When the clock frequency is operating at
320MHz, the DSP arrays could achieve a peak MAC performance of 655.36 GOPS.

\subsection{Block Floating-Point Normalization Module}
\label{subsec:bfp}

\begin{figure}
\setlength{\abovecaptionskip}{0cm}
\setlength{\belowcaptionskip}{1pt}
\centering
\includegraphics[width=0.41\textwidth, scale=0.35,keepaspectratio]{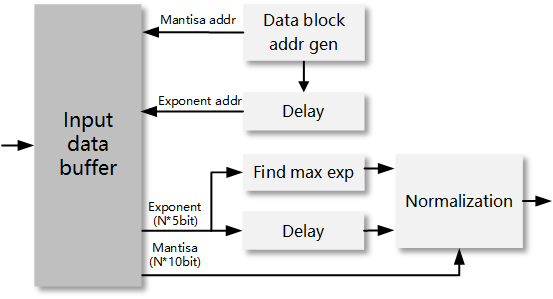}
\caption{The architecture of normalization module.} 
\label{figure:norm}
\vspace{-0.4cm} 
\end{figure}

This design adopts a half-precision floating-point (FP16) representation in storage to maintain relatively high accuracy. However, in MAC computation, block floating-point (BFP) is adopted. A normalization process as shown in Algorithm~\ref{alg:bfp} is required to transform the floating-point data to BFP representation. The normalization module is illustrated in Figure~\ref{figure:norm}. The BFP arithmetic is operated to make an entire block of data sharing a common exponent. This method is possible to maintain a dynamic range similar to floating-point arithmetic while taking advantage of fixed-point computing units. Moreover, the usage of BFP is able to reduce the resource usage of hardware.

\begin{algorithm}[t]
\scriptsize
\caption{BFP normalization algorithm.}
\label{alg:bfp}
\begin{algorithmic}[0]
\STATE {\textbf{Input:} Floating-point number block $\mathbf{X}$}
\STATE {\textbf{Output:} Block floating-point block $\mathbf{X}_{BFP}$}
\STATE {For a block $\mathbf{X}$ containing $N$ floating-point numbers:}
\STATE {\qquad$\mathbf{X} =(x_1,...,x_i,...,x_N)$}
\STATE {\qquad \quad$ = (m_1\times2^{e_1},...,m_i\times2^{e_i},...,m_N\times2^{e_N})$}
\STATE {Find the maximum exponent:}
\STATE {\qquad$\xi_X = \mathbf{max}(e_i),   i\in\{1,2,...N\}$}
\STATE {Normalization:}
    \STATE {\qquad $\mathbf{For}i \gets 1 \mathbf{to} N$}
        \STATE{\qquad\qquad$d_i = \xi_X - e_i$}
        \STATE{\qquad\qquad$m_{bi} = m_i >> d_i$} 
    \STATE {\qquad $\mathbf{EndFor}$}
\STATE {The normalized BFP block:}

\STATE {\qquad\ $\mathbf{X}_{BFP} = (x_{b1},...,x_{bi},...,x_{bN})$}
\STATE {\qquad \quad$ = (m_{b1},...,m_{bi},...,m_{bN})\times2^{\xi_X}$ }

\end{algorithmic}
\end{algorithm}

\addtolength{\topmargin}{0.06in}
\section{Hardware Architecture Optimization}
\label{sec:hardware}

In this section, several methods are presented to improve the efficiency of the hardware architecture. First, parallel skills and image segmentation techniques are applied to parallelize the design. Then, an accuracy maintenance technique during partial summation is proposed.

\subsection{Parallel Techniques}
\label{subsec:paral_tech}

The parallelism is explored efficiently in this work in three aspects: 1) Multiple input/output channels together with 2D MAC arrays guarantee the most fundamental parallel computation; 2) Two sets of input and output buffer facilitate a ping-pong operation. The next round of input data can be pre-loaded during calculation on the current data set. Furthermore, the computing process for the entire convolution is fully pipelined. 3) In the proposed design, the computations of feature extraction, feature fusion and upsample are separated. That means if they are fed with different inputs, the computing processes are independent of each other. Therefore, a multi-threading scheme is proposed to invoke the three modules simultaneously with different inputs. This method could maximize the hardware utilization rate and increase the throughput of the system.

\subsection{Image Segmentation Technique}
\label{subsec:image_seg}

In the practical deployment of the STD system, the input images might be of various sizes. Resizing the image may affect the detection accuracy. In order to reduce the affection from size, the proposed system is designed to support random height images with a width up to 4096. To better fit these types of images into STD computation, a row-wise segmentation technique is utilized in this system. Multiple rows from different input channels are loaded and computed in each round until the entire feature map is scanned. The number of rows and input channels engaged in each round of calculation are dynamically changed according to the current size of feature maps, which can make sure a suitable amount of data is fed into the buffer, thus balance between data loading time and computation time.

If the width of an image exceeds the limit of 4096 pixels while the height
does not, the system would detect this situation and transpose the
image. The proposed system is designed to dynamically support transposed image
computation by transposing the corresponding weight kernels and modifying the
convolution and upsample mode. In the end, the output would be recovered
through an inverse transposition.

Another advantage of using the row-wise segmentation technique is that data reorganization of the input/output feature maps and intermediate results are not required anymore, which reduces plenty of latency. Furthermore, as the row-wise method does not change the storage structure of the image in memory, it can facilitate the DDR burst mode and enhance memory access efficiency.

\subsection{Accuracy Maintenance in Partial Summation}
\label{subsec:acc}

\begin{figure}
\centering
\includegraphics[width=3in, scale=0.35,keepaspectratio]{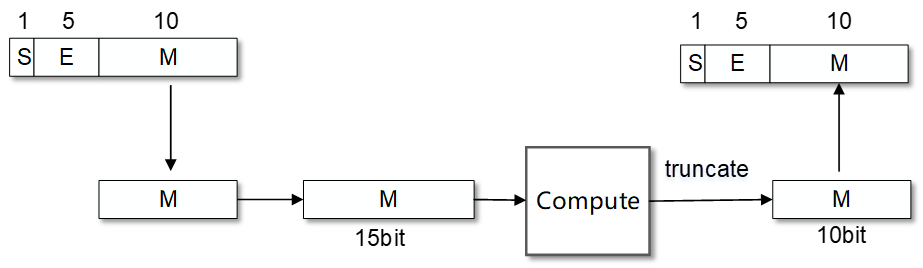}
\caption{The accuracy maintenance procedure that used in partial summation.} 
\label{figure:accuracy}
\vspace{-0.4cm} 
\end{figure}

The model of the STD algorithm is trained by using a single precision
floating-point (i.e. FP32) format. The usage of half-precision
floating-point (i.e. FP16) representation during the inference process will
inevitably introduce a loss of precision. By carrying out an analysis on the
inference process, we found that the partial sum accumulation in the convolution
layer contributes the most to the accuracy loss. This is because the number
of data for summing is enormous, making the loss in each summation
accumulate many times. The results require higher precision than that the
10-bit mantissa in FP16 can provide. 

To address this issue, an accuracy maintenance approach is proposed, and the procedure is illustrated in Figure~\ref{figure:accuracy}. In this approach, the length of mantissa in FP16 is expanded from 10-bit to 15-bit during the partial sum accumulation. The final summation results are truncated back to 10-bit to recover a standard FP16 representation. 
The large dynamic range of 15-bit avoids certain accuracy loss during  accumulation process.

\section{Experimental Results and Comparisons}
\label{sec:exp_result}

\begin{table}
\setlength{\abovecaptionskip}{0cm}
\setlength{\belowcaptionskip}{1pt}
\centering
\caption{Configurations comparison on the GPU and FPGA platforms}
\label{tab:hw_cp}
\begin{tabular}{ ccc }
\hline
~                 & GPU         & FPGA \\
\hline
Type              & NVIDIA Tesla M40  & Xilinx KU115 \\
Memory            & 12G GDDR5     & 16G DDR4 \\
Memory Bandwidth  & 288 GB/s      & 18 GB/s \\
Peak FLOPS        & 7 TFlops      & 2.68 TFlops \\
Max Power         & 250W        & 65W \\
\hline
\end{tabular}
\vspace{-0.4cm} 
\end{table}

The proposed heterogeneous system is implemented using CPU and FPGA. The
hardware architecture is implemented on a Xilinx Kintex UltraScale platform
(XCKU115-FLVA1517-2-E) and the software part is running on an Intel Xeon CPU.
The CPU and FPGA board communicate via a PCIe gen 3$\times$8 interface with 2
channels DDR4 SODIMM.

In terms of the GPU-CPU platform, two main methods are used to increase the
throughput. First, the number of concurrent workers is set to 10. Within each
worker, the batch size is set to 1. According to the experimental results,
this is the optimal setting because further increasing the concurrency would
saturate the GPU memory. Moreover, the input images in the same batch of
different sizes require to be padded and resized to the largest image.
Increasing the batch size would incur additional overhead and reduce
efficiency. Second, the CPU and GPU are configured to work in a pipelined
dataflow to further maximize the throughput. Thanks to the above
optimizations, the GPU utilization rate achieves more than 60\% on average.

\begin{table}
\setlength{\abovecaptionskip}{0cm}
\setlength{\belowcaptionskip}{1pt}
\centering
\caption{Hardware resource utilization of the proposed architecture.}
\label{tab:rsrc}
\begin{tabular}{ cccc }
\hline
~               & Used   & Available  & Utilization (\%) \\
\hline
CLB LUTs        & 286827 & 663360     & 43.24\\
CLB Registers   & 554207 & 1326720    & 41.77\\
DSP48E2         & 2843   & 5520     & 51.69\\
Block RAM Tile  & 999    & 2160     & 46.25\\
CARRY8          & 4638   & 82920    & 5.59\\
\hline
\end{tabular}
\vspace{-0.4cm} 
\end{table}

\subsection{Hardware Resource Utilization}

Table~\ref{tab:rsrc} shows the FPGA resource usage of the proposed
architecture. The hardware utilization rate is designed to be less than $65\%$
because the power supply of the board is from PCIe. The power limitation of
PCIe prevents the design from a higher percentage of hardware usage. As can be
seen from the table that the hardware utilization rate is relatively balanced
in terms of LUT, register, and DSP. Note that the Block RAMs are consumed more
than $60\%$ due to the fact that deploying a ping-pong storage scheme doubles
the number of memories to build input/output buffers.

\begin{figure*}[!t]
 \setlength{\abovecaptionskip}{0cm}
 \setlength{\belowcaptionskip}{1pt}
\centering
\subfloat[]{\includegraphics[width=3in]{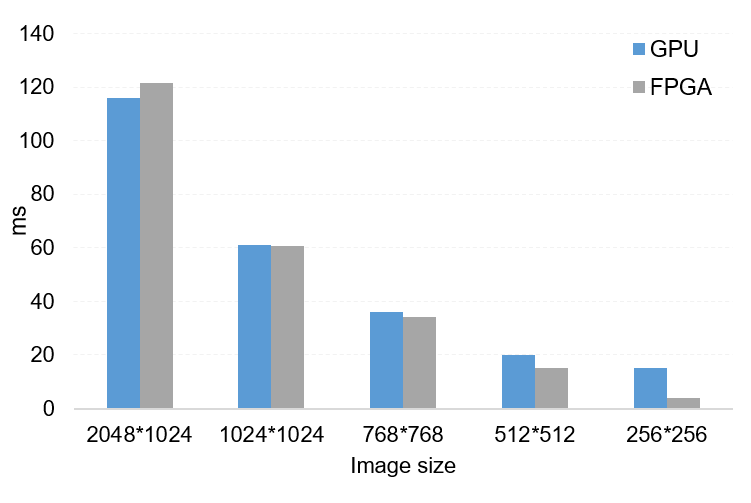}%
\label{figure:lat_cp_r50}}
\hfil
\subfloat[]{\includegraphics[width=3in]{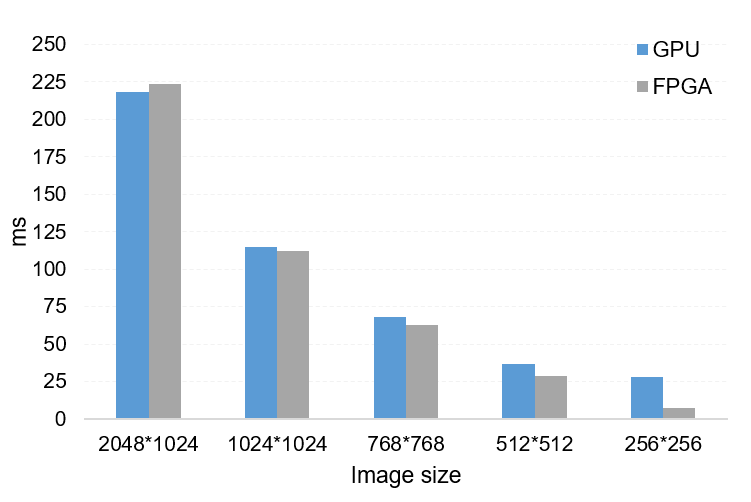}%
\label{figure:lat_cp_vgg}}
\caption{Latency comparison of different size images for ResNet-50 and VGG-16 as feature extractor.(a).ResNet-50,(b).VGG-16}
\label{figure:lat_cp}
\vspace{-0.3cm} 
\end{figure*}

\begin{figure*}[!t]
 \setlength{\abovecaptionskip}{0cm}
 \setlength{\belowcaptionskip}{1pt}
\centering
\subfloat[]{\includegraphics[width=2.6in]{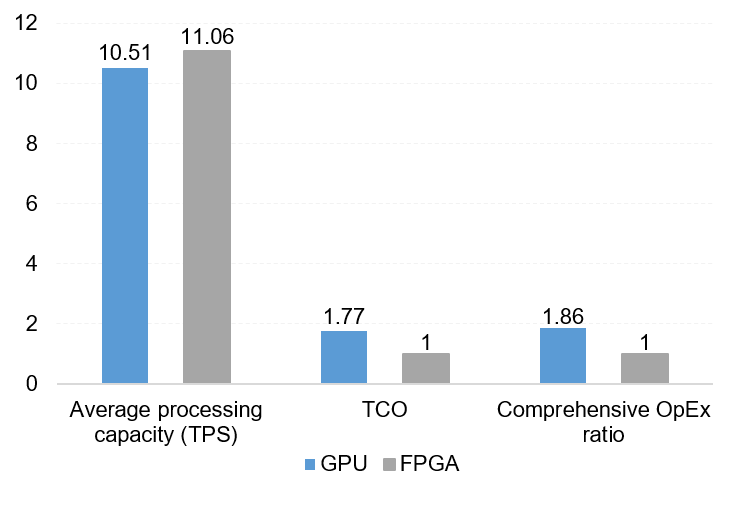}%
\label{figure:opex_cp}}
\hfil
\subfloat[]{\includegraphics[width=2.6in]{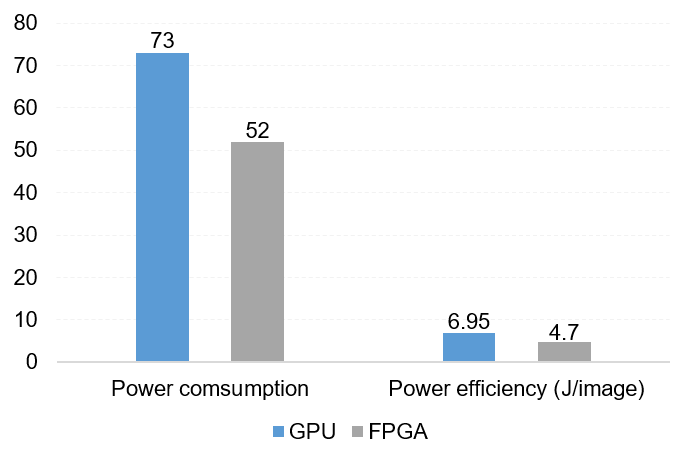}%
\label{figure:power_cp}}
\caption{Comparison between GPU and FPGA in different measurements.(a).TPS and comprehensive OpEx ratio,(b).Power consumption and power efficiency}
\label{figure:opex_cp_power_cp}
\vspace{-0.3cm} 
\end{figure*}

\begin{table*}
\setlength{\abovecaptionskip}{0cm}
\setlength{\belowcaptionskip}{1pt}
\begin{center} 
\begin{scriptsize}
\caption{Comparison of different CNN implementation.}

\label{tab:comp_cnn}
\renewcommand\arraystretch{1.1}
\setlength{\tabcolsep}{0.7mm}
\begin{tabular}{c|c|c|c|c|c|c|c|c|c|c}
\hline
~                       & Ma et al.~\cite{Ma2017}   & \multicolumn{2}{c|}{Cheng et al.~\cite{Cheng2018}}    & Kala et al.~\cite{Kala2019}   & Lian et al.~\cite{LLS+2019}   & \multicolumn{2}{c|}{Xing et al.~\cite{Xing2019}}  & Liang et al.~\cite{Liang2020}     & \multicolumn{2}{c}{Our Work} \\
\hline                                                                                                 
Year                    & 2017                      & \multicolumn{2}{c|}{2018}                             & 2019                          & 2019                          & \multicolumn{2}{c|}{2019}                         & 2020                              & \multicolumn{2}{c}{2021}   \\
\hline                                                                                                                                    
\multirow{2}{*}{FPGA}   & Altera                    & \multicolumn{2}{c|}{Zynq-7000}                        & Virtex-7                      & Virtex-7                      & \multicolumn{2}{c|}{Zynq UltraScale+}             & Zynq UltraScale+                  & \multicolumn{2}{c}{Kintex UltraScale}\\
                        & Arria 10                  & \multicolumn{2}{c|}{XC7Z045}                          & VX690T                        & VX690T                        & \multicolumn{2}{c|}{ZU9}                          & ZCU102                            & \multicolumn{2}{c}{XCKU115} \\
\hline                                                                                                                               
Techonogy               & 20nm                      & \multicolumn{2}{c|}{28nm}                             & 28nm                          & 28nm                          & \multicolumn{2}{c|}{16nm}                         & 16nm                              & \multicolumn{2}{c}{20nm}   \\
\hline                                                                                                                                                                                                                                                                 
Precision               & 16-bit                    & \multicolumn{2}{c|}{8-bit fixed}                      & 16-bit fixed                  & 8-bit BFP                     & \multicolumn{2}{c|}{8-bit fixed}                  & 16-bit fixed                      & \multicolumn{2}{c}{16-bit BFP} \\
\hline                                                                                                                                                                                                                                                                  
Freq (MHz)              & 150                       & \multicolumn{2}{c|}{200}                              & 200                           & 200                           & \multicolumn{2}{c|}{330}                          & 200                               & \multicolumn{2}{c}{320} \\
\hline                                                                                                                                                                                                                                                                
LUTs                    & 128K                      & \multicolumn{2}{c|}{203K}                             & 468K                          & 232K                          & \multicolumn{2}{c|}{118K}                         & 600K                              & \multicolumn{2}{c}{230K}    \\
\hline                                                                                                                                                                                                                                                                  
DSP                     & 1046                      & \multicolumn{2}{c|}{0}                                & 1436                          & 1027                          & \multicolumn{2}{c|}{1542}                         & 2520                              & \multicolumn{2}{c}{2434}    \\
\hline                                                                                                                                                                                                                                                                  
BRAM                    & 2167(20K)                 & \multicolumn{2}{c|}{443}                              & 1465                          & 913                           & \multicolumn{2}{c|}{747}                          & 912                               & \multicolumn{2}{c}{763.5}   \\
\hline                                                                                                                                                                                                                                                                  
Algorithm               & Winograd                  & \multicolumn{2}{c|}{Convention}                       & Winograd                      & Winograd                      & \multicolumn{2}{c|}{Convention}                   & Convention                        & \multicolumn{2}{c}{Winograd \& Convention}\\
\hline                                                                                                                                                                                                                                                                  
CNN power (W)           & /                         & \multicolumn{2}{c|}{10.56}                            & 17.3                          & 9.18                          & \multicolumn{2}{c|}{22.8}                         & 23.6                              & \multicolumn{2}{c}{16.5}   \\
\hline                                                                                                                                                                                                                                                              
Model                   & ResNet-50                 & VGG-16   & ResNet-50                                  & VGG-16                        & VGG-16                        & VGG-16   & ResNet-50                              & VGG-16                            & VGG-16 & ResNet-50      \\
\hline                                                                                                                                                                                                                                                              
Throughput             & \multirow{2}{*}{285.1}    & \multirow{2}{*}{878.1}    & \multirow{2}{*}{804}      & \multirow{2}{*}{407.2}        & \multirow{2}{*}{760.8}        & \multirow{2}{*}{2820}& \multirow{2}{*}{1380}      & \multirow{2}{*}{2479.6}           & \multirow{2}{*}{2866.1} & \multirow{2}{*}{1243.5}   \\
(Average GOPS)          &                           &                           &                           &                               &                               &                      &                            &                                   &\\
\hline                                                                                                                                                                                                                                                                                                                          
Energy efficiency       & \multirow{2}{*}{/}        & \multirow{2}{*}{83.2}     & \multirow{2}{*}{76.1}     & \multirow{2}{*}{23.5}         & \multirow{2}{*}{82.88}        & \multirow{2}{*}{123.7}& \multirow{2}{*}{60.5}     & \multirow{2}{*}{105.4}            & \multirow{2}{*}{173.4} & \multirow{2}{*}{75.2} \\
(GOPS/W)                &                           &                           &                           &                               &                               &                       &                           &                                   &\\

\hline
\end{tabular}
\end{scriptsize}
\end{center}
\vspace{-0.48cm} 
\end{table*}

\subsection{Performance Evaluation}

In this section, we compare the proposed FPGA-CPU heterogeneous platform with GPU-CPU platforms. The detailed configurations of the two platforms are shown in Table~\ref{tab:hw_cp}. 

First, the comparison of latency is conducted. The latency is defined as the average runtime consumption per image. Two candidates of feature extractor network are implemented with ResNet-50 and VGG-16 (without Fully Connected layers). The experimental results of these two platforms are compared and shown in Figure~\ref{figure:lat_cp_r50} and Figure~\ref{figure:lat_cp_vgg}. Overall, the latency performance of GPU and FPGA is closed except for the image of size 256$\times$256. The FPGA outperforms GPU for images of sizes smaller than 1024$\times$1024, while GPU has the advantage in large size images. 

After careful analysis and comparison on throughput and precision, ResNet-50 is selected as the network architecture in the actual deployment. Note that all the following performance comparisons are conducted based on this architecture.

To make fair comparisons, a public dataset is selected in the performance evaluation. The benchmark dataset is the LSVT dataset in ICDAR2019 Robust Reading Competition~\cite{sun2020chinese}. The test dataset is computed by using FPGA-CPU and GPU-CPU systems respectively. The optimal concurrent thread number is set (20 for FPGA-CPU and 10 for GPU-CPU) to explore their best performance. The processing capability in terms of Transactions Per Second (TPS) of two platforms is compared in Figure~\ref{figure:opex_cp}. It is shown that the throughput of FPGA is higher than the GPU platform by 5.2\%. 

The total cost of ownership (TCO) and comprehensive Operating Expense
(OpEx)\footnote{$TCO=\frac{Purchasing~cost +
Maintenance~cost}{Estimated~operating~month}$ and
$OpEx=\frac{TCO}{Througput}$. Due to commercial confidentiality, only the
ratios of FPGA and GPU are given in the text} of the two platforms are compared and shown in Figure~\ref{figure:opex_cp}. The TCO reflects the
operating cost and maintenance cost. It is an important index during
commercial deployment. The TCO ratio of FPGA to GPU in our practice is 1:1.77,
which means that deploying the GPU-CPU system requires 77\% more expense than
the FPGA-CPU counterpart. OpEx indexes are calculated by dividing the TCO by
throughput. Deploying the proposed FPGA-CPU system is able to reduce the OpEx
by approximately 46\%.

In terms of power consumption, the average power of GPU is measured by NVIDIA System Management Interface (vidia-smi), and this index of FPGA board is measured by a digital power meter. The power efficiency (i.e. average power consumed by processing one image, J/image) is calculated and shown in Figure~\ref{figure:power_cp}. Due to the low power consumption and high throughput of the proposed design, it only consumes 4.7J to process one image on average, which is a 32\% decrease compared with the GPU counterpart.

Recently published state-of-the-art CNN accelerators are
thoroughly compared with the proposed FCN architecture, as
shown in Table~\ref{tab:comp_cnn}. The selected objects of
comparison are dedicated implementations for Resnet-50 and VGG-16.
Note that it may be unfair to compare throughput and energy
efficiencies if the implementation devices have different
technologies. However, even though the proposed work uses 20nm
technology FPGA, the average GOPS and GOPS/W indices outperform
the results that using 16nm technology FPGA \cite{Xing2019}~\cite{Liang2020}. This is due to the advantage of our optimizations on
dataflow, adoption of Winograd transform, and multiple parallel
techniques. 

These evaluations show that the proposed design would be very attractive in real commercial products in terms of throughput, operating cost and power efficiency.

\subsection{Precision Results Comparisons}

\begin{table}
\setlength{\abovecaptionskip}{0cm}
\setlength{\belowcaptionskip}{1pt}
\begin{center} 
\caption{Precision results comparisons between GPU and FPGA.}
\label{tab:pc_cp}
\renewcommand\arraystretch{1.1}
\setlength{\tabcolsep}{4.5mm}{
\begin{tabular}{c|c|c|c}
\hline
\multicolumn{4}{c}{ICDAR2019 LSVT Dataset} \\
\hline

Platform    & GPU       & FPGA      & Difference        \\
\hline
Precision   & 84.45\%   & 84.27\%   & -0.21\%           \\
\hline
Recall      & 78.37\%   & 77.69\%   & -0.87\%           \\
\hline
F-measure   & 81.30\%   & 80.85\%   & -0.55\%           \\

\hline
\end{tabular}}
\end{center} 
\vspace{-0.48cm} 
\end{table}

The precise evaluations of FPGA and GPU implementations are also compared by
using ICDAR2019 LSVT dataset, conducted from three perspectives: precision, recall, and f-measure. As shown in Table~\ref{tab:pc_cp}, the results from FPGA are slightly lower than the GPU counterparts, with the largest discrepancy of $0.87\%$. Nevertheless, the precision loss is acceptable to meet the detection system requirements. The f-measure reflects the comprehensive performance, while the decrease of f-measure for the benchmark dataset is only $0.55\%$. After the calculation of a large number of convolution layers, the accuracy loss is able to maintain within an acceptable range, thanks to our accuracy maintenance approach.

\section{Conclusion}
\label{sec:cls}

In this work, a flexible hardware architecture for the instance segmentation based STD algorithm is proposed. The hardware architecture can be dynamically organized through microcode to support different types of FCNs, with the assistance of developed automation software tools. The computing units are fine-grained and optimized, while multiple parallel techniques are exploited to improve efficiency. The implementation results show the proposed design achieves a comparable computing capability, better cost efficiency, and better power efficiency when compared with its GPU counterpart. This work is currently deployed in commercial products to provide consumer scene text detection services with a stable performance.

\section*{Acknowledgment}
The author would like to thank the anonymous reviewers for their valuable comments. This work is supported by the Key-Area R\&D Program of Guangdong Province (2020B0101130003), National Natural Science Foundation of China (No. 62002023), the Guangdong Provincial Key Laboratory of Interdisciplinary Research and Application for Data Science, BNU-HKBU United International College (2022B1212010006), Guangdong Higher Education Upgrading Plan (2021-2025) (UIC R0400001-22), BNU-HKBU UIC Research Grant (R202103), Hong Kong Innovation and Technology Commission (ITF Seed Fund ITS/216/19), City University of Hong Kong (Project Grant No. 9440242 and 9678187).

\bibliographystyle{IEEEtran}
\bibliography{reference}

\vfill

\end{document}